\documentclass[12pt]{article}

\usepackage{amsmath}
\usepackage{amssymb}

\usepackage[koi8-r]{inputenc}
\usepackage[russian]{babel}

\titlepage

\usepackage{graphicx}

\begin{document}


\author{$A.V.Zasov^{1,2}, A.M.Cherepashchuk^{1,2} and L.N.Petrochenko^{2}$}
\title{Super-Massive Black Holes and Dark Halos}
\date{\small $^1$ Physical Department of Moscow State University, Moscow, Russia\\
$^2$ Sternberg Astronomical Institute of Moscow State University, Moscow, Russia}


\maketitle

\begin{center}

{\bf Abstract. }{\small The relations between masses of
Super-Massive Black Holes, $M_{bh}$, in galactic nuclei, maximal
rotational velocities, $V_m$, and indicative masses, $M_i$, of
galaxies are studied for galaxies with the available rotation
curves. $M_{bh}$ correlates with both $V_m$ and $M_i$, although
much weaker than with the central velocity dispersion $S_0$.
Masses $M_{bh}$ for early-type galaxies (S0-Sab), which usually
possess more luminous bulges, are, on average, larger than those
for late-type galaxies with similar velocities of rotation. It
enables to conclude that masses of black holes are determined
mainly by the bulge properties, and - much weaker - by dark
haloes of galaxies.}

\end{center}

\section{Introduction}

A discovery of compact objects, most likely Super-Massive Black
Holes (SMBH), in the nuclei of galaxies, yielded the intensive
discussion on the origin and evolution of these unique objects
and the role they take in the evolution of galaxies. To clarify
the problem it is important to know how the mass of a black hole
is related to masses of different galactic components. The masses
of SMBH have been found to correlate, although with great
dispersion, with the luminosity of spheroidal components of
galaxies. This conclusion was made as far back as 90s
\cite{Korm_Rich_95} and confirmed later on: the most massive SMBH
are observed in the cores of lenticular and elliptical galaxies
of high luminosity. This does not specify, however, whether the
SMBH mass is determined by the processes in the nuclear
environment or the above-mentioned relation appears due to the
mechanism of simultaneous formation of galaxies and the SMBH. Both
alternatives are discussed in literature.

In the first case the SMBH mass, $M_{bh}$, grows with time
because of accretion of matter in the gravitational field of
galactic bulge (see for example the discussion in
\cite{Grupe_Mathur_04, Yoo_Mirada_04}). One of the possible
mechanism is the accretion of gas due to the drag force by the
radiation from bulge stars \cite{Kawakatu_Umemura_04,
Umemura_01}. This approach allows the tight correlation to be
accounted for between $M_{bh}$ and the central dispersion of
stellar velocities, $S_0$. The latter indicates, to some extent,
the depth of the potential well where the stars of these
slow-rotating components are moving.

In the second case, when SMBH forms parallel with the galaxy formation, cosmological scenarios are considered, describing gravitational contraction of the initially expanding
medium which included both barionic matter and non-barionic dark matter. The SMBH mass is
specified in this case by the total mass of a galaxy formed in the field of a dark non-baryonic halo rather than by the bulge mass. A mass of halo, in turn, should  be linked with the maximal (asymptotic) rotation velocity of the galaxy (see discussion in \cite{Ferrarese_02}).

An intermediate opportunity could also be realized with both barionic and non-barionic galactic
components involved in the formation of a black hole. Black hole may grow slowly during the lifetime of
the galaxy due to both stellar accretion and non-adiabatic
accretion of the dark matter, achieving the observed mass over the billions of years \cite{Ilyin_et_al_04, Sirota_et_al_04}.

Although the second scenario does not imply a direct  relation between $M_{bh}$ and velocity dispersion $S_0$, a link between these two values may be explained as a consequence of the correlation of $S_0$ with the maximal galactic rotation
velocity, V$_{m}$. This idea was first put forward in \cite{Ferrarese_02} and supported later in
\cite{Pizella_et_al_03, Baes_et_al_03}.

A direct comparison of the mass $M_{bh}$ with the velocity $V_m$ appears to be the most reliable way to
find correlation between these parameters. However there are not so many galaxies which have both the reliable SMBH mass estimation
and a well defined rotation curve. A conclusion upon relation of
$M_{bh}$ with the rotation velocity was based on correlations between $V_m$ and $S_0$, and between
$M_{bh}$ and $S_0$, existing for galaxies of different morphological types \cite{Pizella_et_al_03,
Baes_et_al_03}. Currently, a growing number of disk (spiral and lenticular) galaxies becomes available for which both $M_{bh}$ and $V_m$ are known, what allows us to study the correlation mentioned above directly.

\section{Correlation between the Black Hole Masses and the maximal  velocities of rotation}

Table 1 includes galaxies which have both SMBH mass estimated and rotation curves measured up to radial distances where they reach maximum or plateau. The second column displays morphological type
of galaxies; the third column is the log of central velocity dispersion.  The fourth one contains maximal disk rotation velocities (in km/s), on the logarithmic scale, obtained
for the disc inclination angle which is presented in the next column ($i=0^0$ for the disk face-on orientation). The last three columns contain the references to the source of rotation curve, log of the SMBH masses (in solar units), and the corresponding references. Masses $M_{bh}$ were taken from
\cite{Cher_03, Merloni_et_al_03}, where  different estimates from the literature were brought  together.


\begin{center}
{\bf TABLE 1.  Galaxies with the available rotation curves and SMBH  masses}
\end{center}

\begin{tabular}{|l|l|c|c|c|c|c|c|}
\hline
Galaxy &  Type & $lg S_o$& $lg V_c$ & i & Reference & $lg M_{bh}$ & Reference\\
\hline
Milky Way & - & - & 2.36 & - & [13] & 6.47 & [11]\\
\hline
NGC224 & Sb & 2.27&2.45 & 77 & [13] & 7.52 & [11]\\
\hline
NGC598 & Sc & 1.56&2.01 & 54 & [13] & <3.18 & [11]\\
\hline
NGC1023 & S0 & 2.31&2.40 & 70 & [14] & 7.59 & [11]\\
\hline
NGC1052 & S0 &2.43 &2.28 & 44 & [14] & 8.29 & [12]\\
\hline
NGC1365 & SBb & 2.18&2.44 & 46 & [13] & 7.66 & [12]\\
\hline
NGC2273 & SBa & 2.09&2.34 & 53 & [15] & 7.27 & [12]\\
\hline
NGC2787 & S0-a & 2.29&2.37 & 55 & [14] & 7.59 & [11]\\
\hline
NGC2841 & Sb & 2.31&2.52 & 68 & [13] & 8.42 & [12]\\
\hline
NGC3031 & Sab & 2.21&2.38 & 59 & [13] & 7.80 & [11]\\
\hline
NGC3079 & SBcd &2.16& 2.38 & 90 & [13] & 7.65 & [12]\\
\hline
NGC3115 & E-S0 & 2.41&2.57 & 67 & [14] & 8.95 & [11]\\
\hline
NGC3169 & Sa & 2.22&2.30 & 63 & [16] & 7.91 & [12]\\
\hline
NGC3227 & SBa & 2.12&2.43 & 56 & [17] & 7.59 & [11]\\
\hline
NGC3384 & E-S0 & 2.17&2.39 & 59 & [14] & 7.26 & [11]\\
\hline
NGC3627 & SBb & 2.06&2.30 & 60 & [18] & 7.26 & [12]\\
\hline
NGC3675 & Sb & 2.03&2.35 & 60 & [19] & 7.11 & [12]\\
\hline
NGC3783 & SBab & 2.19& 2.26 & 25 & [20] & 6.97 & [11]\\
\hline
NGC4051 & SBbc & 1.92&2.20 & 49 & [21] & 6.11 & [11]\\
\hline
NGC4151 & SBbc & 2.19&2.18 & 21 & [22] & 7.18 & [11]\\
\hline
NGC4203 & E-S0 & 2.21&2.36 & 29 & [23] & <7.08 & [11]\\
\hline
NGC4258 & SBbc & 2.13&2.32 & 67 & [13] & 7.61 & [11]\\
\hline
NGC4321 & SBbc & 1.93&2.43 & 27 & [13] & 6.80 & [12]\\
\hline
NGC4388 & Sb & 2.06&2.37 & 90 & [24] & 6.80 & [12]\\
\hline
NGC4395 & SBm & - &1.95 & 71 & [25] & <5.04 & [11]\\
\hline
NGC4450 & Sab &2.11 &2.30 & 44 & [26] & 7.30 & [12]\\
\hline
NGC4459 & S0 & 2.24&2.48 & 42 & [27] & 7.81 & [11]\\
\hline
NGC4501 & Sb & 2.21&2.48 & 56 & [26] & 7.90 & [12]\\
\hline
NGC4548 & SBb & 2.16&2.48 & 38 & [28] & 7.40 & [12]\\
\hline
NGC4565 & Sb & 2.13&2.41 & 86 & [13] & 7.70 & [12]\\
\hline
NGC4579 & SBb & 2.19&2.50 & 36 & [26] & 7.85 & [12]\\
\hline
NGC4594 & Sa & 2.40&2.56 & 84 & [29] & 9.04 & [11]\\
\hline
NGC4725 & SBab & 2.12&2.37 & 51 & [25] & 7.49 & [12]\\
\hline
NGC4736 & Sab & 2.06&2.26 & 35 & [13] & 7.30 & [12]\\
\hline
NGC4945 & SBc & 2.11&2.28 & 78 & [13] & 6.04 & [11]\\
\hline
NGC5033 & Sc & 2.12&2.44 & 55 & [13] & 7.30 & [12]\\
\hline
NGC5194 & Sbc & 1.98&2.41 & 20 & [13] & 6.90 & [12]\\
\hline
NGC6500 & Sab & 2.3&2.48 & 38 & [30] & 8.28 & [12]\\
\hline
NGC7469 & Sba & -&2.08 & 48 & [31] & 6.81 & [11]\\
\hline
Circinus & Sb & -&2.18 & 65 & [32] & 6.11 & [11]\\
\hline
IC342 & SBc & 1.89&2.30 & 25 & [13] & <5.70 & [11]\\
\hline

\end{tabular}

\newpage

In most cases the SMBH masses were estimated using the most
reliable method of reverberation mapping. Other methods give
consistent results within a factor of 2-3 \cite{Cher_03}.

Data on measured rotation velocity and velocity dispersion were
searched using the HYPERLEDA catalogues "Kinematic resolved
catalogue of galaxies" and "Central velocity dispersion"
\cite{Hyper}. The maximal rotation velocities in Table 1
correspond to inclination angles assumed in the original papers.
If there were some independent sources of rotation curve, the one
with the most reliable curve was preferred. In the case where
both absorption and emission line measurements were available
only the latter was considered. For some early-type galaxies
rotation velocities were measured using only absorption lines. In
these cases asymptotic drift, that is stellar velocity
dispersion, was taken into account \cite{Neistein_et_al_99}. In
some galaxies a rotation curve passes through a local maximum
within 1 - 2 kpc from to the center after which the curve
flattens of resumes to grow. Such maximum was not considered to
belong to the galactic disk or halo, and, therefore, was ignored
while assessing $V_m$. The galaxies, which display strong
large-scale non-circular gas motion were not included in the
sample.

The maximal rotation velocities of the galaxies, $V_m$, versus BH masses, $M_{bh}$, are shown
in Fig. 1.




\renewcommand{\figurename}{Fig.}
\begin{figure} 
\includegraphics[scale=0.55,angle=270]{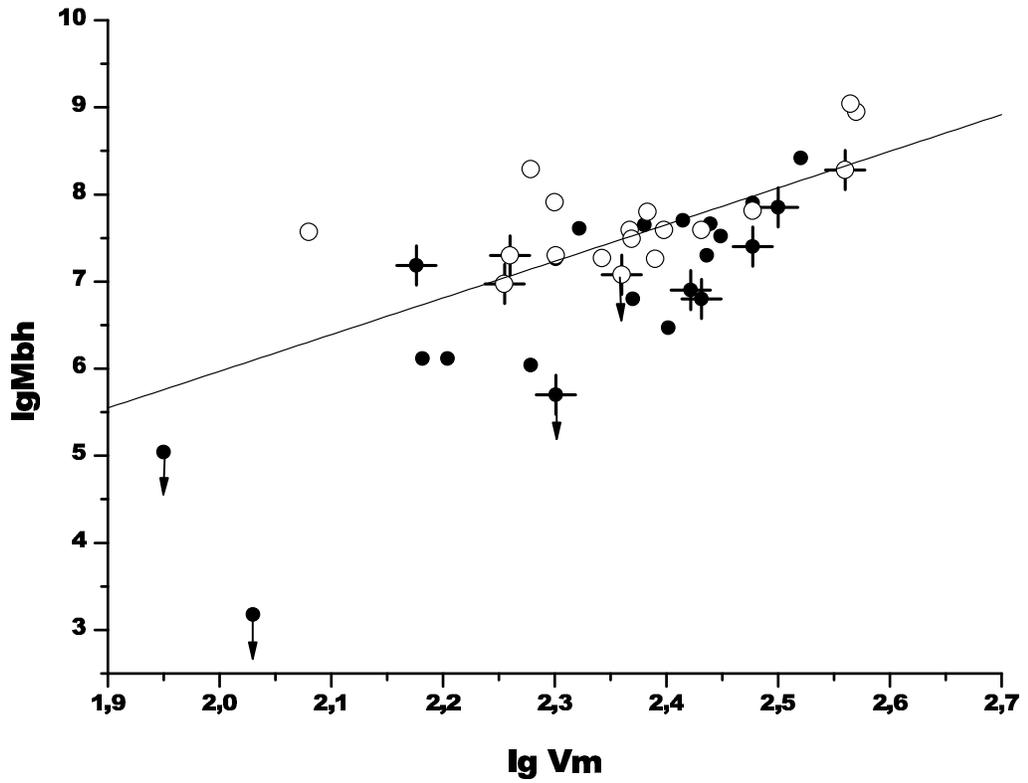}
\caption{Maximal rotation velocities vs SMBH masses diagram.
Non-filled symbols stand for the early-type S0-Sab galaxies,
crosses indicate galaxies with the disk inclination $i<40^o$. The
straight line shows a relation found in \cite{Baes_et_al_03}
using indirect $V_m$ estimates.} \label{Fig_1}
\end{figure}

Non-filled symbols mark early-type galaxies, S0-Sab, possessing the most luminous bulges. The diagram does not show a tight correlation between
the parameters considered (correlation coefficient being k=0.74), although
 slowly
rotating disk galaxies in all cases have low-mass black holes. Two close interacting ga\-la\-xies from the
sample (NGC 3227 and NGC 5194) do not appear aside in the plot. The lowest point  in the diagram belongs to the M33 galaxy for which only upper mass limit of SMBH is known.

Figure 1 shows some difference between galaxies of early and late morphological types.
S0-Sab galaxies have, on the average, more massive black holes as compared to late-type galaxies with similar velocity $V_m$.

The straight line in the diagram represents a relation between the values under comparison that was found in
\cite{Baes_et_al_03} using indirect estimates of $V_m$. The agreement with our data is rather poor,
most points lying below the straight line in the plot.

Figure 2 displays the relation between $M_{bh}$ and the central dispersion of velocities, $S_0$, for
a sample of galaxies  (the data are not available
for some objects).

\begin{figure}
\includegraphics[scale=0.55, angle=270]{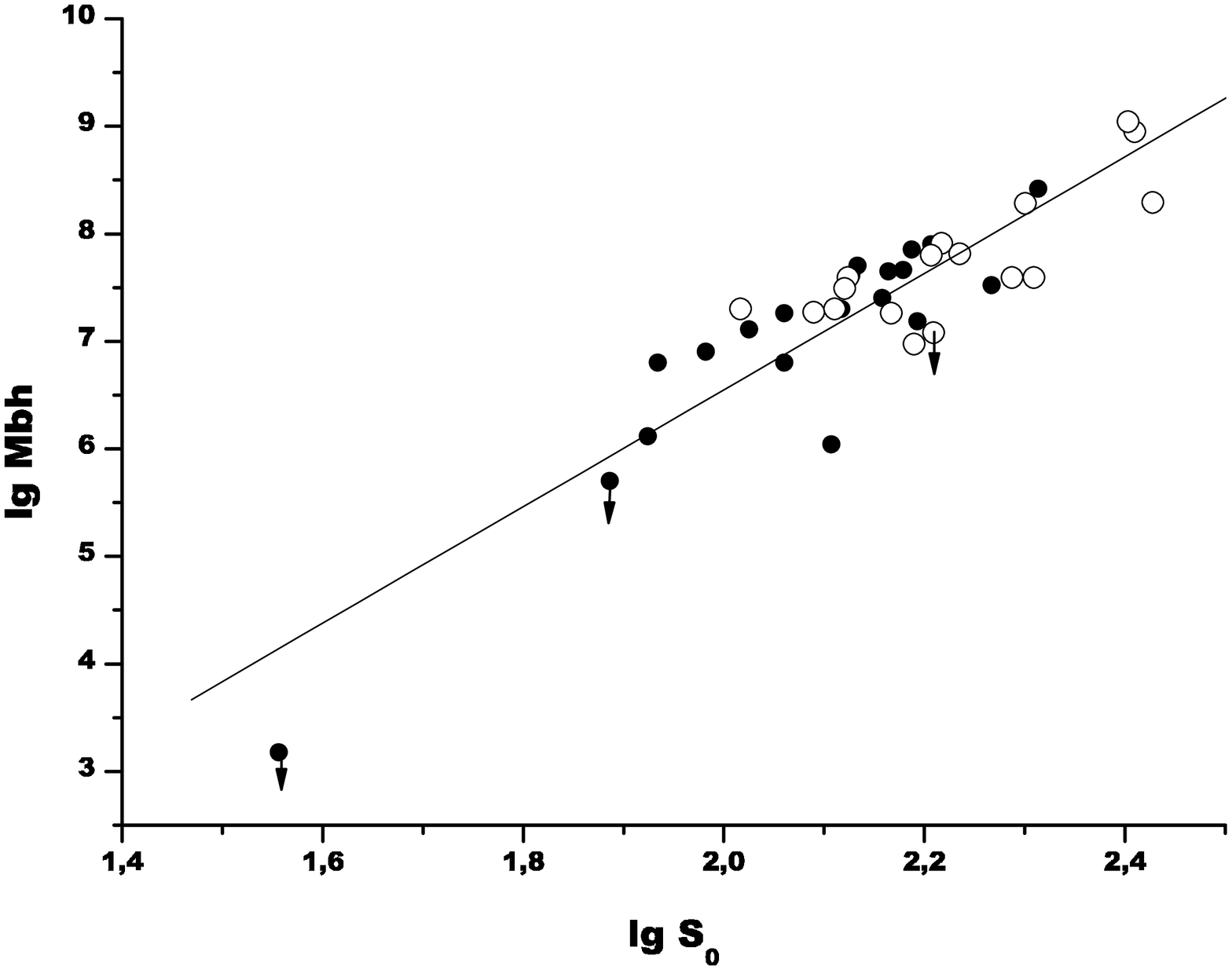}
\caption{Relation between the black hole mass and the central
dispersion of velocities. Non-filled circles stand for the
early-type S0-Sab galaxies.}
\label{Fig_2}
\end{figure}

The correlation between these two parameters is tighter than  between $M_{bh}$ and
$V_m$, the correlation coefficient being equal $k=0.90$ for all galaxies and 0.92 for the spiral galaxies
from the sample. It is noteworthy that both the early-type galaxies (non-filled symbols) and the
late-type ones lie in Figure 2 along the same sequence whereas in Figure 1 the early-type galaxies
have systematically greater SMBH masses in a wide range of rotational velocities.

It is necessary to take into account that the velocities of
rotation are determined with less accuracy than the central
velocity dispersions. The main sources of uncertainty are errors
in the assumed disk orientation angles (the inclination angle, i,
and the position angle of the kinematic major axis, $PA_0$). The
uncertainties they induce are the most essential for nearly
face-on galaxies (the discussion concerning casual and systematic
errors in determining orientation angles of spiral galaxies see
e.g. in \cite{Fridman_04}). For instance, if $i=30^0$, and its
error is about $\Delta i=\pm 10^0$ this error would change the
resulting value of $\lg V_m$ by -0.11 or +0.16 depending on the
error sign. Galaxies with $i<40^0$ are marked in Fig.1 by crosses
as the objects with the least reliable $V_m$ assessed. They do
not seem to be a real source of observational point scattering.

The other source of uncertainty is a complex shape of rotation curves, which in often makes the value of $V_m$ rather ambiguous.
In many cases the observed rotation curve continues to increase or to decrease on a disc periphery, not revealing a smooth plateau (see some examples in Fig.3). Moreover, this behavior is hardly the result of  the measurement errors: as it was shown in [16], the sign and the meaning of velocity gradients at large $R$ correlate with he total luminosities of galaxies.

\begin{figure}
\includegraphics[scale=0.53,angle=270]{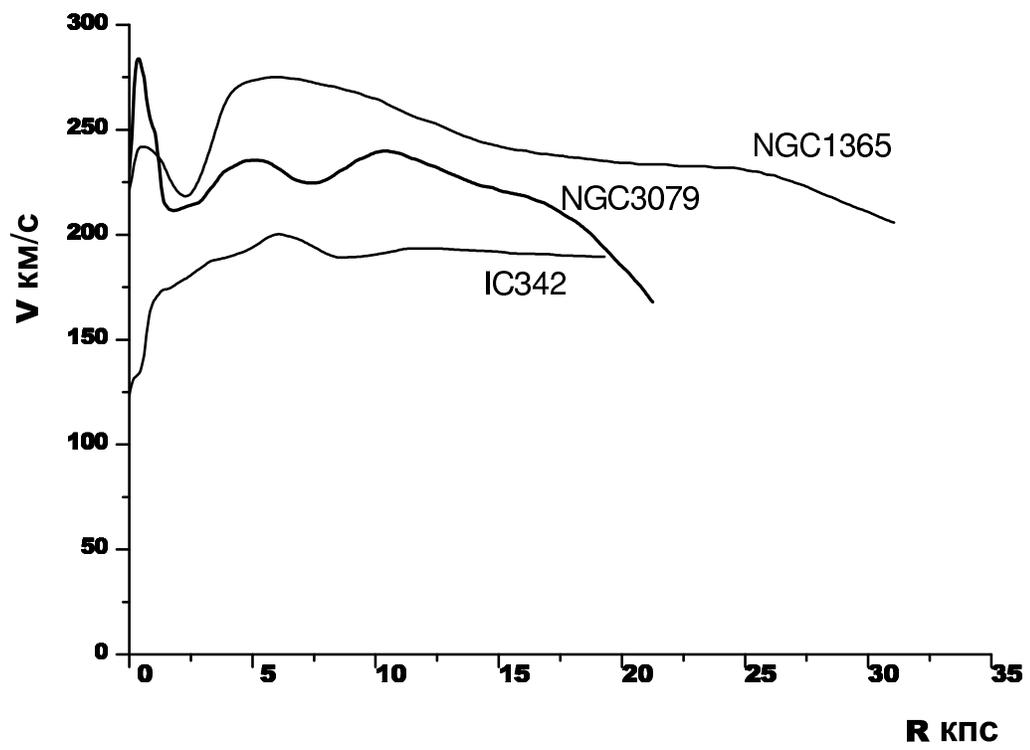}
\caption{Examples of different shapes of rotation curves of
galaxies considered (from Sofue et al.[17]). } \label{Fig_3}
\end{figure}

To reduce the uncertainty when assessing $V_m$, we have extracted from the list of galaxies in Table 1 a subsample of galaxies with the most reliable rotational curves. First of all, these are galaxies entering into the representative
samples of galaxies with
extended rotation curves by
Y.Sofue et al. \cite{Sofue_et_al_99} and S.Casertano, J.H.Gorkom \cite{Casertano_Gorkom_91}. The rotation curves of these galaxies cover at least $0.9 R_{25}$ where
$R_{25}$ is a radius of the galaxy up to the isophote $B$=25 $mag/sq.arcsec$. For the early-type spirals to be included, our subsample contains also five nearest non-interacting S0 galaxies, velocities of rotation of which were determined using the same approach \cite{Neistein_et_al_99}, and Sombrero galaxy (NGC 4594), type Sa, with a very bright bulge. Its rotation curve, obtained from emission line measurements, is given in \cite{Rubin_et_al_85}. Although the rotational curves of these
early-type galaxies are less extended than those of the other galaxies of the subsample, they still extend up to distance $R=(0.3-0.7)R_{25}$, beyond the inner region of a galaxy with a steep velocity gradient.

For the galaxies where the velocity curve falls in the periphery, both
the maximal rotation velocity, $V_m$, within the rotation curve and the rotation velocity at the
maximal distance from the center, $V(R_{max})$, were considered separately. Since a relative contribution of the
dark halo into the mass of the galaxy increases with $R$, the velocity
$V(R_{max})$ is expected to be bound up with the halo properties more tightly than $V_m$. There is
an indirect evidence in favor of this suggestion, namely, the Tully-Fisher relation between the luminosity
of  galaxies and their rotation velocity becomes more tight if to use the velocity of far-from-center regions \cite{Verheijen_01}.

Figure 4 displays the SMBH mass versus $V_m$ and $V(R_{max})$
(non-filled circles). It follows from this diagram that the
relation between $M_{bh}$ and the rotation velocity for the
subsample of galaxies remains very "loose". The use of
$V(R_{max})$ instead of $V_m$ makes scattering of points in the
plot even greater (k=0.75 against k=0.82). It confirms indirectly
that $M_{bh}$ depends very weakly on the halo mass. The
dispersion of points in the diagram was not found to depend on
the inclination angle of the galactic disk either.

\begin{figure}
\includegraphics[scale=0.5,angle=270]{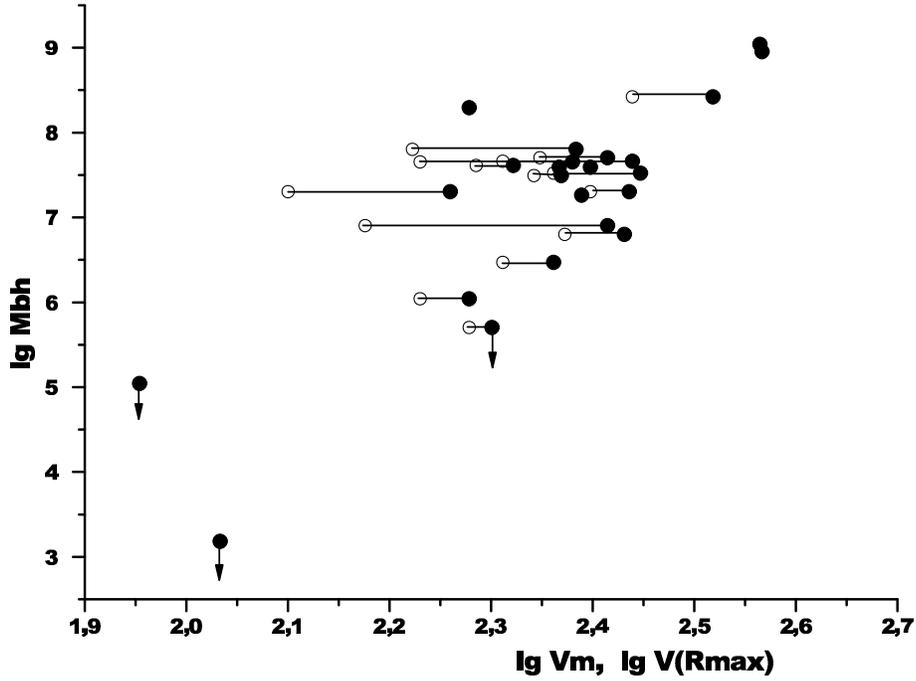}
 \caption{Rotation velocity vs black hole
mass diagram for a subsample of galaxies with the most reliable
rotation curves. For those galaxies where the maximal rotation
velocity, $V_m$, and the velocity at the edge of the rotation
curve, $V(R_{max})$, are noticeably different, both values are
plotted and connected by a horizontal line.}
\label{Fig_4}
\end{figure}

A similar conclusion
can be drawn while comparing $M_{bh}$ with an indicative mass of a galaxy,
$M_i=V_m^2 R_{25}/G$, which is close to a total mass of a galaxy containing within the optical radius of a disc. Note that the dark halo mass of spiral galaxies usually dominates over luminous mass within this limit
(see, for example, \cite{Salucci_01, Zasov_et_al_02, Zasov_et_al_04} and references therein).
Figure 4 displays the $M_i - M_{bh}$ plot for the galaxies of the subsample we consider. Large
scattering of the points in the diagram demonstrates that the correlation between the black hole mass
and the total galaxy mass remains rather weak.

\begin{figure}
\includegraphics[scale=0.5,angle=270]{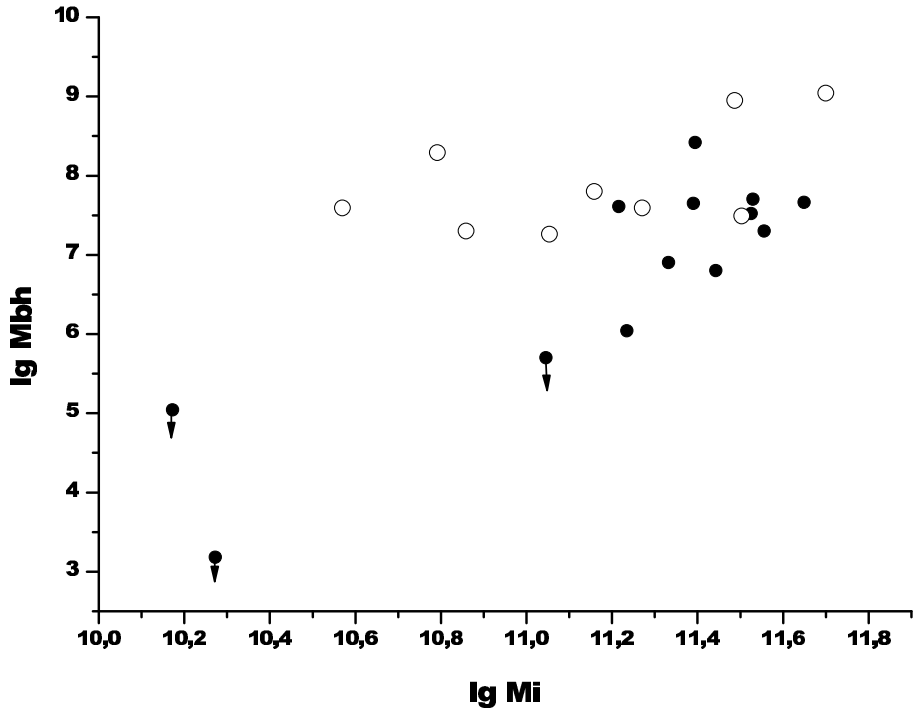}
\caption{Indicative mass of a galaxy vs black hole mass diagram.
Non-filled circles stand for the early-type
 S0-Sab galaxies.}
\label{Fig_5}
\end{figure}

\section{Conclusions}

Correlations between SMBH masses and parameters depending on the
dark halo mass ($V_m$, $V(R_{max})$, or $M_i$) which are
determined directly from the rotation curves, are much looser
than their correlation with the central velocity dispersion.
Moreover, early-type disk galaxies with a higher relative
luminosity of their bulge  usually possess more massive black
holes than late-type galaxies for the same rotational velocities
(at least in the interval 150-250 km/s), although these galaxies
follow the same relation as the late-type spirals if the
correlation between $M_{bh}$ and the central velocity dispersion
is considered (Fig. 2). These facts confirm the presence of a
strong link between a black hole and a bulge, and rather weak
link with halo.

As Fig. 1 shows, a relation between the rotational velocity and the black hole mass
becomes especially loose, if exist at all, for slow rotating galaxies (although $M_{bh}$ remains to be correlated
with their central velocity dispersion. This property is to be taken into account while considering the process of SMBH formation.

A tight connection between SMBH and a bulge is the argument in favor of a scenario according to
which the central black hole is growing due to accretion processes during the period of whole life
of a galaxy rather
than it forms together with the galaxy.

Different processes of black hole growth, discussing in the literature, include the accretion of gas, stars, dark matter (see f.e. [6, 7, 8] and references therein) or even the accretion black holes formed in globular clusters [39]. It is possible however that the initial masses of primeval black holes in galactic nuclei depend on the process of a bulge formation in a potential well (cusp) created by a massive dark halo. In this scenario a halo may indirectly
influence the process of mass growing and its present day value \cite{Ilyin_et_al_04, Sirota_et_al_04}.

To clarify how the SMBH mass grows with time, the rate of accretion onto the black hole is to be
estimated in a direct way. Recently such estimates were carried out (within certain simplifications) for a number of galaxies,
from the intensities of [OIII] line in galactic nuclei
\cite{Heckman_et_al_04}. The results obtained indicate that characteristic mass growth time for a SMBH
is rather different, depending on whether the black hole is a low-mass or a massive one, being longer
for the latter. The growth time of SMBH seems to be comparable with the
growth time of the surrounding stellar bulge. The suggestion does not conflict with the conclusion that
current SMBH masses and their growth depends on the processes occurring now (or occurred in the past) in the bulges of galaxies.

Authors thank A.D. Chernin for stimulating discussion and A.V.Tutukov for the valuable remarks.

\bigskip

The work is supported by the Russian Federal Principal Scientific and Technical Program (contract
40.022.1.1.1101) and RFBR grant 04-02-16518.

\newpage

\bigskip
\renewcommand{\refname}{References}


\begin{thebibliography}{99}
\bibitem{Korm_Rich_95}
Kormendy J., Richstone D. Ann. Rev. Astron. Astrophys. 1995. V.33. P.581.
\bibitem{Grupe_Mathur_04}
Grupe J., Mathur S. Astrophys. J. 2004. V.606. P.L41.
\bibitem{Yoo_Mirada_04}
Yoo J., Mirada-Escude J. Astrophys. J. 2004. In press (astro-ph 0406217).
\bibitem{Kawakatu_Umemura_04}
Kawakatu N., Umemura M. Astrophys. J. 2004. V.601. P.L21.
\bibitem{Umemura_01}
Umemura M. Astrophys. J. 2001. V.560. P.L29.
\bibitem{Ferrarese_02}
Ferrarese L. Astrophys. J. 2002. V.578. P.90.
\bibitem{Ilyin_et_al_04}
Ilyin A.S., Zybin K.P., Gurevich A.V. Zh. Eksp. Teor. Fiz. 2004. V.98, P.5 (astro-ph 0306490).
\bibitem{Sirota_et_al_04}
Sirota V.A., Ilyin A.S., Zybin K.P., Gurevich A.V. Astro-ph 0403023.
\bibitem{Pizella_et_al_03}
Pizella A., Corsini E.M., Vega Beltran J.C., et al. Mem. SAIt. 2003. V.74. P.504.
\bibitem{Baes_et_al_03}
Baes M., Buyle P., Hau G.K.T., Dejonghe H. MNRAS. 2003. V.341. P.L44.
\bibitem{Cher_03}
Cherepashchuk A.M. Uspekhi Phys. Nauk. 2003. V.173. P.345.
\bibitem{Merloni_et_al_03}
Merloni A., Heinz S., di Matteo T., et al. MNRAS. 2003. V.345. P.257.
\bibitem{Hyper}
HYPERLEDA: http://www.sai.msu.ru/hypercat
\bibitem{Neistein_et_al_99}
Neistein E., Maoz D., Rix H.-W., Tonry J.N. Astron.J. 1999. V.117. P.2666.
\bibitem{Fridman_04}
Fridman A.M., Afanasiev V.L., Dodonov S.N., et al. Accepted in MNRAS. 2004.
\bibitem{Casertano_Gorkom_91}
Casertano S., Gorkom J.H. Astron.J. 1991. V.101. P.1231.
\bibitem{Sofue_et_al_99}
Sofue Y., Tutui Y., Honma M., et al. Astrophys.J. 1999. V.523. P.136.
\bibitem{Rubin_et_al_85}
Rubin V.C., Burstein D., Ford W.K., Thonnard N. Astrophys.J. 1985. V.289. P.81.
\bibitem{Verheijen_01}
Verheijen M.A.W. Astrophys.J. 2001. V.563. P.694.
\bibitem{Salucci_01}
Salucci P. MNRAS. 2001. V.320. P.L1.
\bibitem{Zasov_et_al_02}
Zasov A.V., Bizyaev D.V., Makarov D.I., Tyurina N.V. Pis'ma v Astron.Zh. 2002. V.28. P.599.
\bibitem{Zasov_et_al_04}
Zasov A.V., Khoperskov A.V., Tyurina N.V. Pis'ma v Astron.Zh. 2004. In press.
\bibitem{Tut}
Tutukov A.V. Astron.Rev. 2004, submitted.
\bibitem{Heckman_et_al_04}
Heckman T.M., Kauffmann G., Brinchmann J., et al. Astrophys.J. 2004. V.613. P.109.
\end{thebibliography}
\end{document}